\documentclass[12pt]{iopart}
\usepackage{graphicx}

\begin{document}

\title{Drift mode accelerometry for spaceborne gravity measurements}

\author{John W. Conklin\dag\  
\footnote[3]{To
whom correspondence should be addressed (jwconklin@ufl.edu)}
}

\address{\dag\ Department of Mechanical and Aerospace Engineering,
  University of Florida, Gainesville, FL 32611}

\begin{abstract}
A drift mode accelerometer is a precision device that overcomes the much of the
acceleration noise and readout dynamic range limitations of
traditional electrostatic accelerometers.
It has the potential of achieving acceleration noise performance of drag-free
systems over a restricted frequency band without the need for external
drag-free control or spacecraft propulsion.
Like traditional accelerometers, the drift mode accelerometer contains
a high-density test mass surrounded by an electrode housing, which can control
and sense all six degrees of freedom of the test mass.
Unlike traditional accelerometers, the suspension system is operated with
a low duty cycle so that the limiting suspension force noise only acts
over brief, known time intervals, which can be neglected in the data analysis.
The readout is performed using a laser interferometer which is immune
to the dynamic range limitations of even the best voltage references
typically
used to determine the inertial acceleration of electrostatic accelerometers.
The drift mode accelerometer is related to the like-named operational
mode of the LISA Pathfinder spacecraft, which will be used to estimate the
acceleration noise associated with the LISA Pathfinder front end electronics.
This paper describes operation of such a device, develops models for its
performance with respect to satellite geodesy and gravitational wave
astrophysics applications, and discusses methods for testing its performance
using torsion pendula in the laboratory and the LISA Pathfinder mission
in space.
\end{abstract}



\maketitle

\section{Introduction}

Precise measurement of inertial acceleration is vital to many space-borne
gravitational science missions, including satellite geodesy \cite{GRACE},
fundamental
physics experiments \cite{microscope, prl} and gravitational wave observation
\cite{LISA}.
The most precise accelerometers manufactured to date are the electrostatic
accelerometers produced by ONERA, which are capable of measuring
spacecraft acceleration relative the the inertial frame  to $\sim 10^{-11}
\ \mathrm{m/sec^2 Hz^{1/2}}$ from roughly 1 mHz to 1 Hz
\cite{touboul2012}.
These accelerometers have been used for Low-low Satellite-to-satellite tracking
missions including GRACE \cite{GRACE}
and for gravity gradiometer missions such as GOCE \cite{GOCE}.

These instruments are comprised of an internal free-floating metallic
test mass that is surrounded by an electrode housing.
The electrodes on the internal surface of the housing both sense the test
mass' position capacitively and actuate it via electrostatic forces.
The position measurement is used to drive the electrostatic suspension system
to keep the test mass centered in its housing.
The inertial acceleration of the spacecraft is proportional to the suspension
force applied to the test mass to keep it centered.

Electrostatic accelerometers are limited by two inter-related factors: 1)
suspension force noise and 2) acceleration measurement noise.
Both are ultimately
related to the stability of voltage references, where the current
state of the art is $\sim 2 \times 10^{-6}$ \cite{GOCEaccel}.
For the application of Earth geodesy the low frequency acceleration of a
low Earth orbiting satellite can be as high as
$\sim 10^{-5} \ \mathrm{m/sec^2}$.
Therefore the resulting acceleration
noise on the test mass due to the suspension system is at least
$2 \times 10^{-11} \ \mathrm{m/sec^2}$.
Since the applied suspension force \emph{is} the acceleration measurement,
the acceleration measurement noise would be on this same order.
To improve accelerometers significantly beyond the $10^{-11}
\ \mathrm{m/sec^2}$ level,
the suspension force noise must be removed \emph{and} the sensor used to
measure acceleration must be changed.

Drag-free technology, conceived of in the 1960's \cite{Lange, debra},
has been the
most promising approach to breaking through these acceleration noise limits.
Two drag-free approaches have been demonstrated on three separate missions.
The first is an ``accelerometer-mode'' drag-free, where an electrostatic
accelerometer is used as the primary sensor and a propulsion system is used
to counter the drag-force acting on the satellite so that the nominal
test mass suspension force is reduced.
The spacecraft acceleration measurement is still limited by voltage reference
stability, but the nominal voltage applied to the housing electrodes
is reduced, therefore the electrostatic force noise is also reduced.
Both Gravity Probe B \cite{prl}
and the GOCE \cite{GOCE} missions operated in accelerometer-mode
drag-free.
Using this approach Gravity Probe B achieved an acceleration
noise of $4 \times 10^{-11} \ \mathrm{m/sec^2 Hz^{1/2}}$ \cite{bencze}
and GOCE achieved a differential acceleration noise measurement between test
masses
accurate to $\sim 10^{-12} \ \mathrm{m/sec^2 Hz^{1/2}}$ in the 1 mHz to 1 Hz
frequency band \cite{GOCEaccel}.

The other drag-free operating mode is 'true' drag-free, where the suspension
force is turned completely off, at least in one degree of freedom.
Triad I with its DISturbance COmpensation System (DISCOS)
operated in this manor \cite{discos1, discos2},
as will the Laser Interferometer Space Antenna (LISA)
in the future.
A fundamental difference between accelerometers and true drag-free is that
the basic measurement for a true drag-free system is displacement
variations, instead of acceleration variations.
Of course one can always convert displacement to acceleration and vice-versa.

A drift-mode accelerometer (DMA) as defined here is a traditional electrostatic
accelerometer where the test mass suspension force is operated with a low
duty cycle.
Larger suspension forces are used, but over a much shorter period of time
so that the average suspension force is the same as that of a traditional
accelerometer.
By switching the suspension system on and off with a constant frequency and
low duty cycle ($< 0.1$), the suspension system force noise is
restricted to known, short intervals, which repeat with a frequency chosen to be
above the science frequencies of interest.

Cycling the suspension system eliminates suspension force noise while the
suspension system is off,
but there still is the problem of precisely
measuring the inertial acceleration of the satellite in the presence of a
large zero-frequency acceleration.
Here, laser interferometry provides the solution.
In most upcoming precision gravity missions, the measurement of interest
is the relative displacement (or acceleration) between two or more
inertially fixed test masses.
GRACE Follow-on, GRACE-II, and LISA \cite{LISA}
are all examples.
In all of these missions, the laser interferometer system already exists and is
used to measure range variations between spacecraft.
If an interferometer is used to also measure distance variations
between a reference point on the spacecraft and the test mass, then this
measurement can be used to estimate the inertial acceleration of the
spacecraft, assuming that the test mass can be treated as inertially fixed
over the short interval when the suspension system is off.
Second order finite differencing provides the simplest method.
Although other approaches discussed in this paper
can provide more accurate estimates.

Laser interferometers have been demonstrated with extremely large dynamic
range.
The LISA Interferometric Measurement System for example can measure
pm variations over 1000 sec between spacecraft that have relative
velocities of 10 m/sec.
This represents a dynamic range of $10^{22}$.

The name drift mode is taken from an operating mode of LISA Pathfinder (LPF)
\cite{LPF, LPFdriftmode}.
LPF contains two free-floating test masses.
The spacecraft can only fly drag-free about one of them (naturally) and,
therefore, the other test mass must be suspended against the gravity
gradient (and other) forces which act upon it.
In order to assess the acceleration noise associated with the suspension
electronics, the drift mode was conceived.
The suspension system is turned on and off with a low duty cycle (1 sec on and
200 sec off).
In between ``kicks'' the test mass follows approximate parabolic
trajectories, when measured relative the the other test mass.
These parabolas are fit to second order polynomials and the fit residuals
are used to calculate variations in the differential acceleration between the
two test masses.
Since the goal of the drift mode for LPF is simply to determine the
acceleration noise on the test masses due to the actuation electronics,
the time between kicks was chosen to be relatively long (200 sec).
The interferometer data during the kicks is discarded and replaced with
a model of the acceleration noise that makes various assumptions
about the nature of the noise \cite{LPFdriftmode}.
In contrast, for the DMA we wish to make no assumptions about the
inertial acceleration of the satellite and therefore, we choose a kicking
frequency that lies above the science signal of interest.

\section{Acceleration noise}

The acceleration noise budget for precision accelerometers typically
contains roughly 30 known acceleration noise terms.
The acceleration noise budgets provided here are based on
models used for the LISA mission \cite{schumaker2003, gerardi2014}.
These individual noise terms can be categorized by their physical nature
such as
magnetic, electrical, thermal, Brownian, etc.
In this paper the individual noise terms are grouped into four main
categories: (1) gap-dependent, (2) gap-independent, (3) actuation and sensing,
and (4) stiffness.
Gap-dependent acceleration noise sources are those which fundamentally depend
of the size of the gap between the test mass and its housing.
For GRACE-like accelerometers, these gaps are on the order of
$\sim 100 \ \mathrm{\mu m}$.
Gap-dependent noise sources are typically the dominant source of acceleration
noise and are the reason why the LISA gravitational reference sensors,
which were originally based on the ONERA accelerometers use relatively large
gaps of 4 mm along the sensitive direction.
Gap-independent acceleration noise comprises all bulk test mass
forces, including magnetic and gravitational noise, as well as surfaces
forces which do not depend on the gap size.

The third type of acceleration noise is actuation and measurement noise.
As discussed in the introduction section, for electrostatic accelerometers
both actuation and measurement
noise is ultimately due to the instability of voltage references.
Measurement noise represents the noise associated with making the
acceleration measurement.
For electrostatic accelerometers, the actuation force applied to the
test mass to keep it centered in its housing \emph{is} the
acceleration measurement.

\begin{figure}
  \centering
  \includegraphics[width = 10 cm]{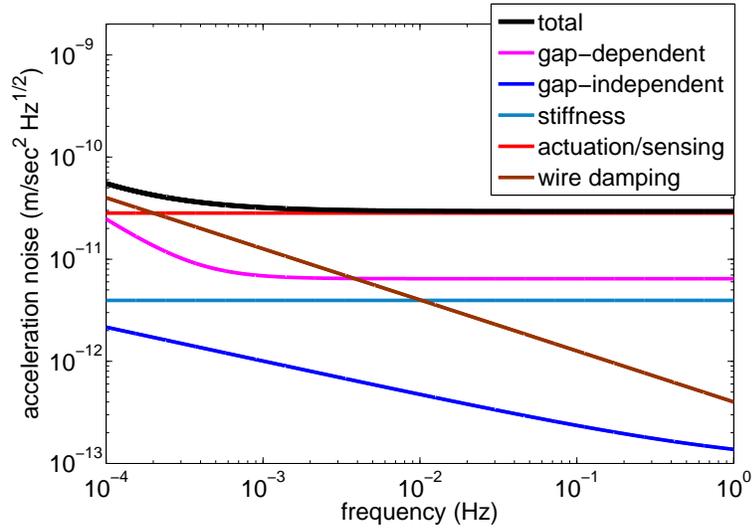}
  \caption{Approximate acceleration noise budget for a GRACE-like
    accelerometer. \label{fig:GRACEaccel}}
\end{figure}

Figure, \ref{fig:GRACEaccel} provides a rough breakdown of the contributions
to acceleration noise for GRACE like accelerometers.
Table \ref{tab:GRACEparam} provides the key parameters used to produce
Figure \ref{fig:GRACEaccel} following the methodology outlined in
\cite{gerardi2014}.
The acceleration noise budget for GRACE-like accelerometers is limited by
measurement and actuation noise as previously stated.
If we assume a relative stability of the voltage reference of $2\times10^{-6}$,
and a nominal drag-induced acceleration of $10^{-5}
\ \mathrm{m/sec^2 Hz^{1/2}}$,
then the suspension force noise acting on the test mass is $2\times10^{-11}
\ \mathrm{m/sec^2 Hz^{1/2}}$.
Here, the measurement noise is assumed to be the same.

\begin{table}
\caption{\label{tab:GRACEparam} Basic design parameters of a GRACE-like
electrostatic accelerometer and a candidate DMA for Earth geodesy
following the methodology of \cite{gerardi2014}.}
\lineup
\begin{tabular}{@{}lll}
\br
Parameter & GRACE-like accelerometer \cite{touboul2012}
  & DMA for Earth geodesy \\
\mr

Mass of TM & 72 g & 243 g \\

TM/housing gap & 175 $\mu$m & 1 mm \\

Surface area of TM & $4 \times 10^{-4} \ \mathrm{m^2}$ &
  $9 \times 10^{-4} \ \mathrm{m^2}$\\

Charge control & Au wire & UV photoemission \\

Surface area of spacecraft & $1 \ \mathrm{m^2}$ & $1 \ \mathrm{m^2}$ \\

Mass of spacecraft & 100 kg & 100 kg \\

Magnetic susceptibility of TM & $2 \times 10^{-6}$ & $2 \times 10^{-6}$ \\

TM stray voltage & 100 mV & 100 mV \\

Max. TM charge &  $1 \times 10^{7}$ e &
  $1 \times 10^{7}$ e\\

Max. dc magnetic field & 50 $\mu$T & 50 $\mu$T \\

Max. magnetic field fluctuation & 1 $\mathrm{ \mu T / Hz^{1/2} }$
  & 1 $\mathrm{ \mu T / Hz^{1/2} }$ \\

Max. magnetic field gradient & 10 $\mu$T/m & 10 $\mu$T/m \\

Max. field gradient fluctuations & 0.25 $ \mathrm{ \mu T/m Hz^{1/2} }$
  & 0.25 $ \mathrm{ \mu T/m Hz^{1/2} }$ \\

Pressure inside housing & 10 $\mu$ Pa & 10 $\mu$ Pa \\

Temperature difference
  & $10^{-2}$(1mHz/$f$)$^{(1/3)}$ $\mathrm{ K / Hz^{1/2} }$
  & $10^{-2}$(1mHz/$f$)$^{(1/3)}$ $\mathrm{ K / Hz^{1/2} }$ \\
fluctuations across housing & & \\

\br
\end{tabular}
\end{table}

In Figure \ref{fig:GRACEaccel} there is one additional noise source related
to controlling the buildup of charge on the test mass.
These accelerometers use an extremely thin
$(\sim 10 \ \mathrm{\mu m}$ diameter)
gold fiber to electrically ground the test mass to its electrode housing.
This wire contributes a thermal force noise on the test mass with a $1/f^{1/2}$
spectrum.

\begin{figure}
  \centering
  \includegraphics[width = 10 cm]{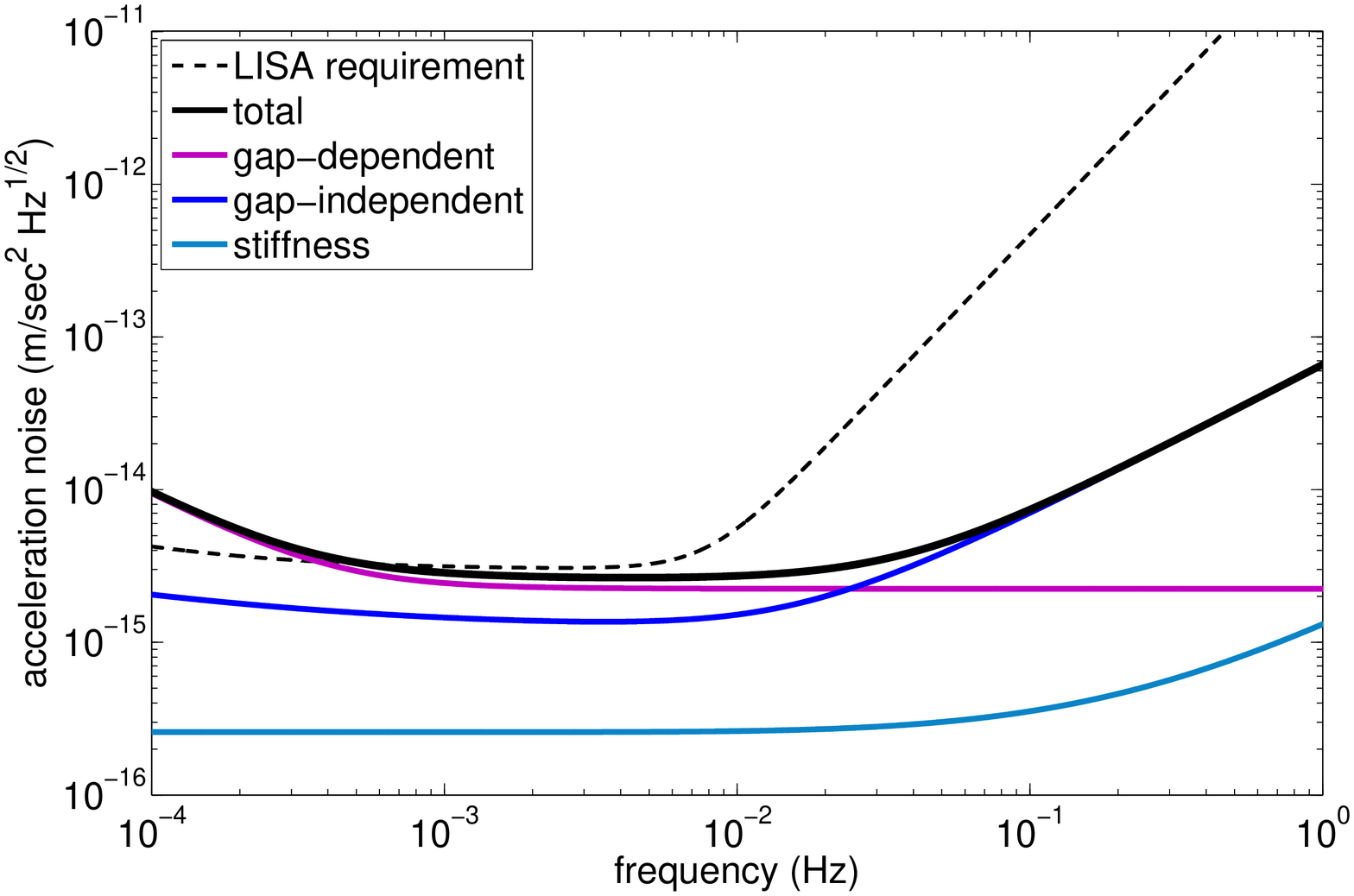}
  \caption{Acceleration noise budget for LISA. \label{fig:LISAaccel}}
\end{figure}

Figure \ref{fig:LISAaccel} shows the approximate acceleration noise
budget for LISA, with individual noise terms grouped as before.
This budget follows that of \cite{schumaker2003, gerardi2014}.
Also shown in Figure \ref{fig:LISAaccel} is the requirement for LISA,
$3 \times 10^{-15} \ \mathrm{m/sec^2Hz^{1/2}}$ from roughly 0.1 - 10 mHz.
Since LISA is operated 'true' drag-free mode
there is no test mass actuation and therefore no associated acceleration
noise.
Two other factors that greatly improve the performance of the LISA
GRS relative to GRACE are larger gaps (10$\times$ that of GRACE),
and a non-contact charge control system, based on photoemission using
UV light \cite{shaul2008}.
The second difference eliminates the thermal noise of the gold fiber
used in the GRACE accelerometers.

\subsection{A drift mode accelerometer model}

\begin{figure}
  \centering
  \includegraphics[width = 14 cm]{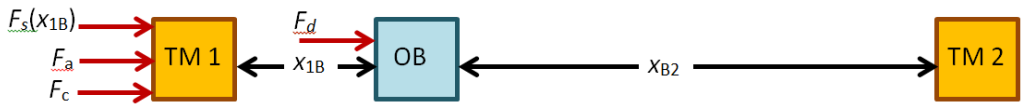}
  \caption{One-dimensional model used to estimate the performance
  of a drift mode accelerometer. \label{fig:model}}
\end{figure}

In order to estimate the acceleration noise performance of a drift mode
accelerometer
the following model, depicted in Figure \ref{fig:model} is used.
In this model two test masses are widely separated.
Test mass 2 (TM 2) is assumed to be inertially fixed for simplicity.
The goal of the DMA is measure the inertial acceleration of the spacecraft which
houses test mass 1 (TM 1).
Measurement of the spacecraft's motion relative the TM 1 is made relative
to a optical bench (OB), which is assumed to contain two laser interferometers.
The first measures the position of TM 1 relative to OB, $x_{1B}$, and the
second measures OB relative to TM 2, $x_{B2}$.

Three forces act on the TM 2: control forces denoted $F_c$, position-dependent
forces (stiffness forces) denoted $F_s(x_{1B})$, and all other disturbance
forces
$F_{a}$.
The force $F_a$ consists of both gap-dependent and gap-independent
forces described
above.
The disturbance force applied to the spacecraft is denoted $F_d$ and
is largely due to atmospheric drag in the case of Earth geodesy missions
and solar radiation pressure for deep space gravitational wave and
other fundamental physics missions.

A simple PID control law is implemented to keep TM 1 centered in its housing
$(x_{1B} = 0)$.
The controller was cycled on and off with a periodicity of $T_{\mathrm{kick}}$
seconds with a duty cycle of 0.1.
The disturbance force applied to the spacecraft is mission dependent
and therefore Earth geodesy and gravitational wave applications
were analyzed separately. The results are described below.

\subsection{DMA for Earth geodesy}

\begin{figure}
  \centering
  \includegraphics[width = 10 cm]{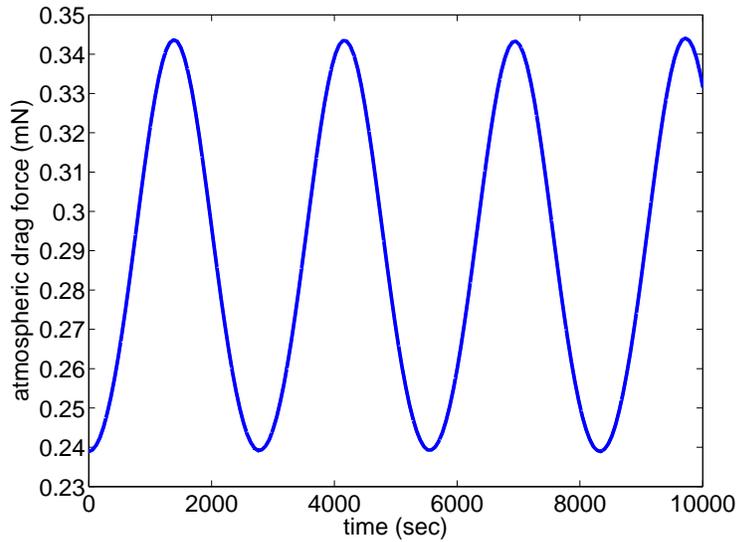}
  \caption{Time-history of the atmospheric drag force acting on
  a 1 $\mathrm{m^2}$ satellite in a 400 km circular polar orbit.
  \label{fig:dragVtime}}
\end{figure}

\begin{figure}
  \centering
  \includegraphics[width = 10 cm]{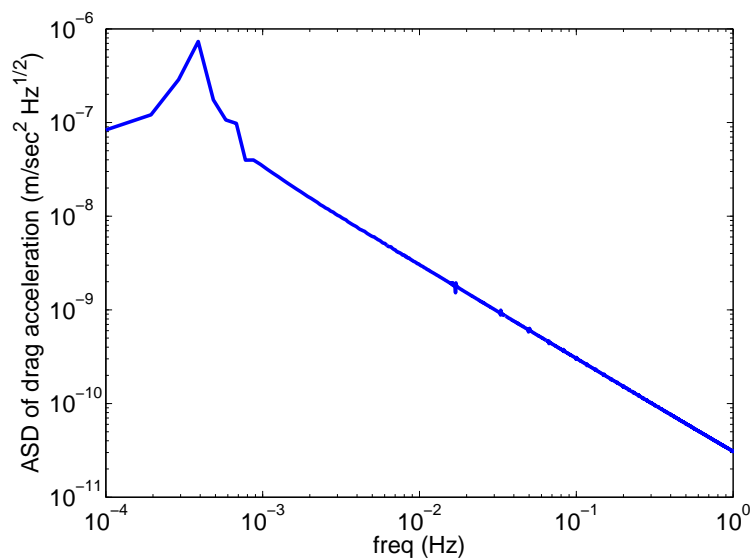}
  \caption{Amplitude spectral density of atmospheric drag acceleration
    acting on a 100 kg satellite in a 400 km circular polar orbit.
  \label{fig:dragAccel}}
\end{figure}

For the simulation of a low Earth orbiting satellite, atmospheric drag
is calculated using the NRLMSISE-00 empirical atmospheric model
acting on the spacecraft in a 400 km circular polar orbit.
The spacecraft mass and cross sectional area is 100 kg and 1 $\mathrm{m^2}$
respectively, and its coefficient of drag is $C_D = 1$.
Figure, \ref{fig:dragVtime} shows the time history of the atmospheric drag
force acting on the satellite and Figure \ref{fig:dragAccel} shows the 
spectrum of the corresponding drag acceleration.
The main variations in the drag force occur at twice the orbital frequency.

\begin{figure}
  \centering
  \includegraphics[width = 10 cm]{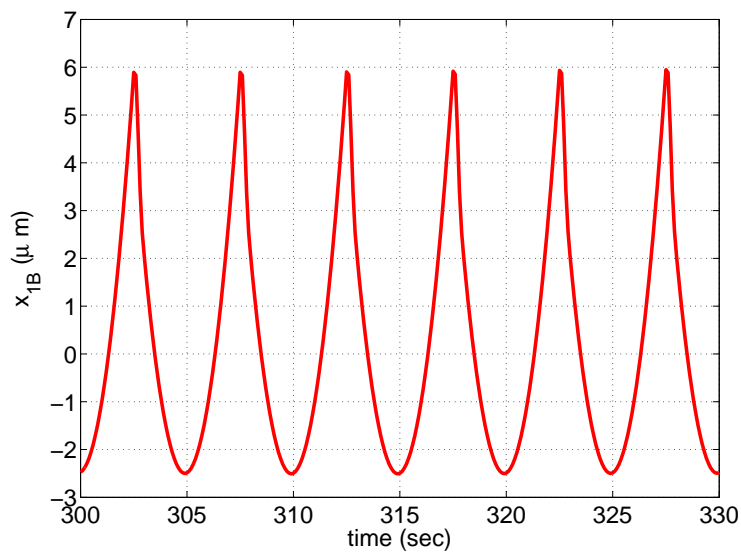}
  \caption{Time history of test mass position along the sensitive axis
    for a drift mode accelerometer in low Earth orbit. \label{fig:juggle}}
\end{figure}

An actuation cycling period of $T_{\mathrm{kick}} = 5 \ \mathrm{sec}$ is
chosen and the
resulting position time history of TM 1 relative to the spacecraft
($x_{1B}$) is shown in Figure \ref{fig:juggle}.
Parabolic trajectories with a frequency of 0.2 Hz and an amplitude of
$\sim 4 \
\mathrm{\mu m}$ are apparent.

\begin{figure}
  \centering
  \includegraphics[width = 10 cm]{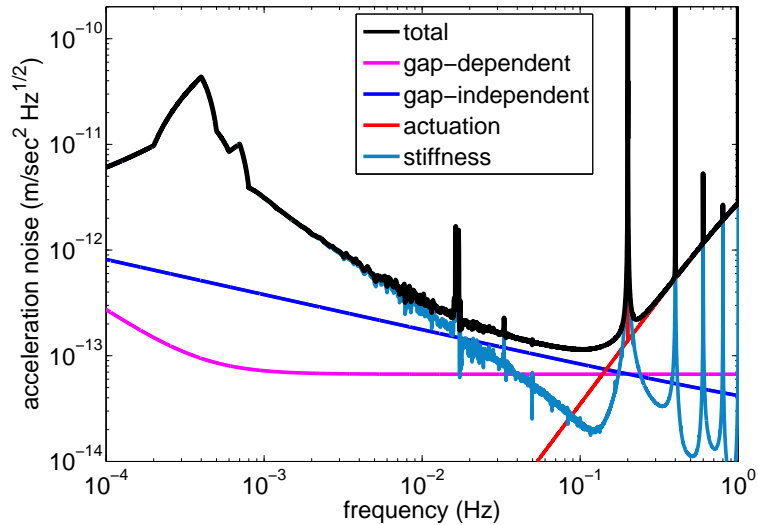}
  \caption{Acceleration noise for a drift mode accelerometer for
    Earth geodesy. \label{fig:GRACEaccelDM}}
\end{figure}

To calculate the acceleration noise performance of a drift mode
accelerometer designed for Earth geodesy, a candidate instrument
is chosen with basic properties listed in Table \ref{tab:GRACEparam}.
All key design features of the DMA are kept the same as that of the
GRACE-like accelerometer, expect the size and mass of the test mass are
increased to 30 cm and 243 g respectively, and a
TM-to-housing gap size of 1 mm is used.
In addition, it is assumed that the gold wire used for test mass charge control
is eliminated and replaced with a charge control system utilizing
UV photoemission.
Figure \ref{fig:GRACEaccelDM} shows the estimated performance of such an
instrument.
Gap-dependent and gap-independent acceleration noise terms are calculated
as before.
Actuation noise is modeled simply as the spectrum of maximum applied
acceleration
multiplied by a relative voltage noise of $2 \times 10^{-6}$, with a
0.1 duty cycle a repetition rate of 1/5 sec.
In reality, for a DMA the acceleration data is discarded while the actuation
system is on and spacecraft acceleration is only estimated using the data
when the actuation system is off.
Therefore, if we assume we retrieve one acceleration measurement per actuation
cycle, then the maximum frequency of the acceleration noise spectrum
should be 0.2 Hz.
By comparing Figures \ref{fig:GRACEaccel} and \ref{fig:GRACEaccelDM}
we see that the broadband acceleration noise of
$2\times10^{-11} \ \mathrm{m/sec^2}$ for the traditional accelerometer
is frequency shifted to 0.2 Hz plus harmonics.

As we can see in Figure \ref{fig:GRACEaccelDM}
the limiting acceleration noise term is stiffness, which
is the coupling of the $\sim 4 \ \mathrm{\mu m}$ motion of the
spacecraft relative to TM 1 and the stiffness, $k = 2 \times 10^{-6} \
\mathrm{sec^{-2}}$.
Most of the stiffness related acceleration noise contribution occurs
at the suspension cycling frequency of 0.2 Hz and its harmonics.
Contributions at lower frequencies, especially twice the orbital frequency,
are caused by the low frequency
contribution of the atmospheric drag.
As discussed below, if the stiffness $k$ can be be determined through
calibration, then the stiffness-related acceleration noise can be subtracted
in the data analysis.
The resulting acceleration noise of the DMA for Earth geodesy would then
be $\sim 4 \times 10^{-13} \ \mathrm{m/sec^2 Hz^{1/2}}$ around 1 mHz.

\subsection{DMA for gravitational wave observation}

\begin{figure}
  \centering
  \includegraphics[width = 10 cm]{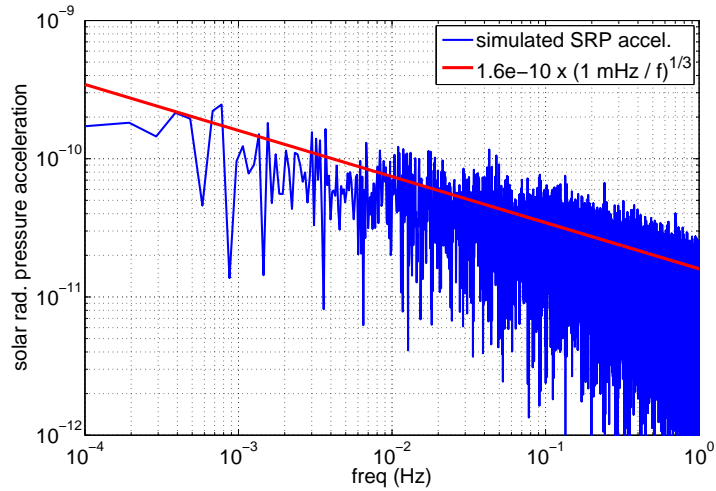}
  \caption{Actual and simulated spectra of the solar radiation pressure
    acceleration acting on a LISA-like
    spacecraft a distance of 1 AU from the Sun \label{fig:SRP}}
\end{figure}

In order to assess the performance of the DMA for with respect to
gravitational wave observation, the geometry and other properties of the
accelerometer were assumed to be the same as the LISA Pathfinder
gravitational reference sensor.
The LPF GRS is a 2 kg, 46 mm, Au/Pt cube, with 4 mm gaps along the
sensitive axis between the test mass and its electrode housing.
For a 500 kg spacecraft at a distance of 1 AU from the Sun, the zero-frequency
spacecraft acceleration due to solar radiation pressure is,
\begin{eqnarray}
  a_0^{\mathrm{SRP}} & = & P_\odot \, A / M =
    ( 4.6 \times 10^{-6} \ \mathrm{N/m^2} )(4 \ \mathrm{m^2})
    / (500 \ \mathrm{kg}) \\
    & = & 4 \times 10^{-8} \ \mathrm{m/sec^2}. \nonumber
\end{eqnarray}
The high frequency solar radiation pressure, taken from \cite{schumaker2003}
is, $a^{\mathrm{SRP}} \approx 1.6 \times 10^{-10} \left( 1 \ \mathrm{mHz} /
f \right)^{1/3} \ \mathrm{m/sec^2 \, Hz^{1/2}}$.
Solar radiation pressure acceleration amplitude spectral density and the
spectrum of the numerically simulated acceleration are shown in Figure
\ref{fig:SRP}.

\begin{figure}
  \centering
  \includegraphics[width = 10 cm]{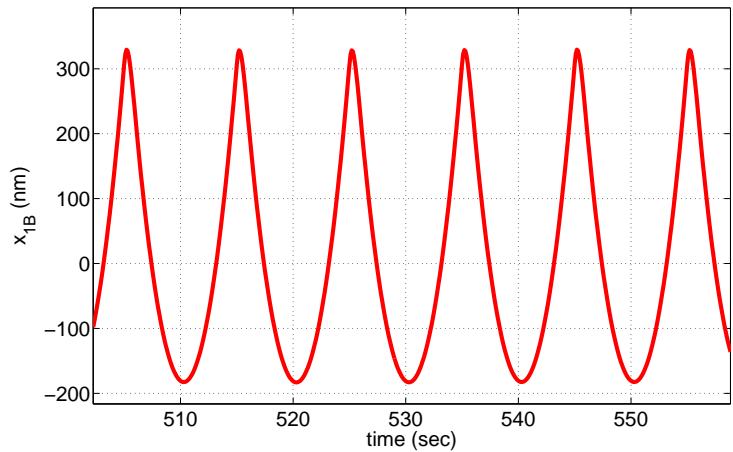}
  \caption{Time history of test mass position along the sensitive axis
    for a drift mode accelerometer in a LISA-like spacecraft in 
    heliocentric orbit. \label{fig:LISAjuggle}}
\end{figure}

If the suspension system is operated with a repetition rate of 0.1 Hz
and a duty cycle of 0.1, the resulting parabolic motion of TM 1
relative to the optical bench has an amplitude on the order of
250 $\mathrm{n m}$ as shown in Figure \ref{fig:LISAjuggle}.

\begin{figure}
  \centering
  \includegraphics[width = 10 cm]{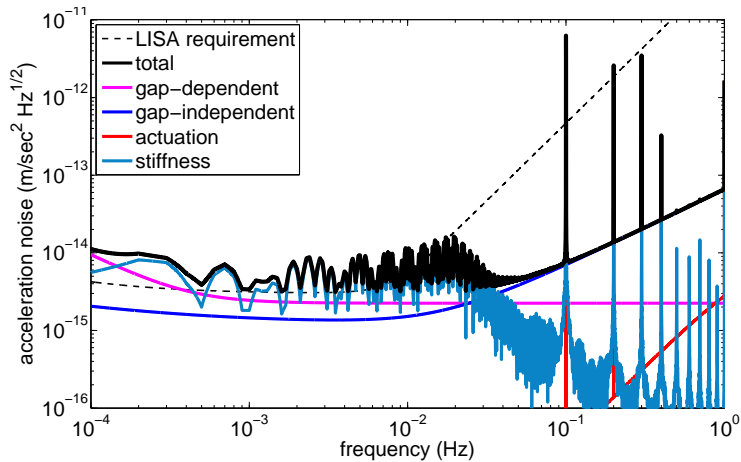}
  \caption{Acceleration noise for a drift mode accelerometer for
    gravitational wave astrophysics. \label{fig:LISAaccelDM}}
\end{figure}

Figure \ref{fig:LISAaccelDM} shows the resulting performance of the DMA
with respect to gravitational wave observation.
As is the case for the Earth geodesy DMA,
the limiting acceleration noise term is stiffness, which
again is the coupling of the $\sim 250 \ \mathrm{nm}$ motion of the
spacecraft relative to TM 1 and the stiffness, $k = 10^{-7} \
\mathrm{sec^{-2}}$.
Most of the stiffness related acceleration noise contribution occurs
at the suspension cycling frequency of 0.1 Hz and its harmonics.
Contributions at lower frequencies are caused by the low frequency
contribution of the solar radiation pressure acceleration noise acting
on the satellite.
As with the geodesy application, if the stiffness can be determined through
calibration, then the stiffness-related acceleration noise can be subtracted
in the data analysis.

In both applications there exist at least
two acceleration noise sources that may be calibrated and removed in the
data analysis.
They are stiffness (position-dependent) forces and actuation
cross-coupling forces.
Both are not fundamentally limiting (e.g. unavoidable quantum mechanical
effects) and can therefore be calibrated and removed.
However, a rigorous determination of the possible accuracy of such a
calibration must still be determined.

\subsection{Position-dependent noise}

Because of the increased motion of the test mass relative to its housing
the stiffness related force noise is
much larger than
that of the GRACE accelerometers or the drag-free LISA GRS.
These position dependent forces do not represent a fundamental limit to the
performance of the DMA.
If stiffness $k$ can be determined through calibration, then the measured
position of TM 1 relative to the spacecraft $x_{1B}$ can be used to
estimate $F_s(x_{1B})$ and subtract it in the data analysis.
Procedures for estimating the stiffness to high precision
have been developed
for LISA Pathfinder \cite{LTPDA2011},
though these techniques rely on measuring the motion of one
test mass relative to another.
Determination of $k$ for the DMA might therefore require
measurement of position of TM 1 relative to TM 2 ($x_{12} = x_{1B} - x_{B2}$),
which is of course readily available.
The stiffness related signal present in $x_{12}$ would be primarily at the
kicking frequency, which is chosen to be above the dominant science signals
of interest.
Therefore, estimation of $k$ from these data should be cleaning separable from
the science signal.

In order for position-dependent acceleration noise to be reduced below the
fundamental limit of $\sim 4 \times 10^{-13} \ \mathrm{m/sec^2 Hz^{1/2}}$
around 1 mHz for the Earth geodesy DMA,
the stiffness $k$ must be determined with a relative accuracy of 0.1 or
an absolute accuracy of $2 \times 10^{-7} \ \mathrm{sec^{-2}}$.
For the gravitational wave DMA the stiffness is much lower
($1\times10^{-7} \ \mathrm{sec^{-2}}$)
because of the increased gap size, the larger test mass,
and the stricter requirements on the
environmental stability of the GRS.
Already, the position-dependent acceleration noise, shown in Figure
\ref{fig:LISAaccelDM}, for gravitational wave observation is near the
fundamental limit.
To drop it below this limit calibration accuracy must be a modest 0.2
relative to $k$ or $5\times10^{-8} \ \mathrm{sec^{-2}}$ absolute.

Of course, increasing the actuation cycling frequency reduces the
spacecraft-to-TM motion and therefore reduces stiffness related acceleration
noise.
One could therefore choose a cycling frequency that is high enough
to reduce the stiffness related acceleration noise to below the fundamental
limit.
However, as we will see in Section \ref{sec:measNoise}, reducing the
actuation cycling frequency dramatically increases the interferometric
acceleration  measurement noise.


\subsection{Actuation cross-coupling}

Two different methods can be used to suspend the test mass in all
rotational degrees of freedom and in all
translational degrees of freedom orthogonal to the sensitive direction.
These degrees of freedom can either be continuously supported or operated in
drift mode just like the sensitive degree of freedom.
All degrees of freedom can be operated in drift mode only if the resulting
motion does not cause loss of performance of the interferometer, which
measures displacement along the sensitive axis.
This generally requires a relatively high cycling frequency, which again results
in a relatively large acceleration measurement noise as discussed
in Section \ref{sec:measNoise}.

If we assume that all degrees of freedom except the degree of freedom along
the sensitive axis are suspended continuously against
the external forces applied to the host spacecraft, then we
must consider the additional
acceleration noise acting in the sensitive direction due to
actuation cross-coupling.
Actuation cross-coupling is the inadvertent forcing of the test mass in the
sensitive direction, which occurs when actuating the test mass in another
degree of freedom due to a small residual coupling $\lambda$.
This cross coupling can be as large as $\lambda = 5\times10^{-3}$ for inertial
sensors like that of LISA.
For both geodesy and gravitational wave applications, this cross coupling
acceleration exceeds the fundamental acceleration noise limit in the sensitive
direction.

If these cross coupling coefficients can be determined, 
then using the known
applied forces in all degrees of freedom, the resulting force
in the sensitive direction can be calculated or eliminated with the
appropriate combination of applied electrode voltages.
Determination of such coefficients has been precisely demonstrated by the
GOCE mission and will also be performed during the LISA Pathfinder
mission \cite{LTPDA2011}.

One technique for determining these cross-coupling coefficients is to
dither the actuation voltages in each of the non-sensitive degrees of freedom
and fit a model of the cross-coupling to the interferometric measurement
of the test mass motion along the sensitive axis.
To roughly determine how well the cross coupling coefficient $\lambda$
can be determined using this approach,
the numerical simulation described above was modified to include a
dither voltage equivalent to a test mass acceleration of
$0.5 \ \mathrm{\mu m/sec^2}$ on a perpendicular axis with a frequency of
10 mHz.
For a $10^4$ sec simulation, the interferometer readout along the sensitive
axis with an assumed measurement noise of $10^{-11} \ \mathrm{m/Hz^{1/2}}$
was capable of estimating $\lambda$ with a relative accuracy of
$5\times10^{-4}$.

Examining Figure \ref{fig:dragAccel} we see that the atmospheric drag
acceleration at 1 mHz is $\sim 3 \times 10^{-8} \ \mathrm{m/sec^2 Hz^{1/2}}$.
If we assume a cross coupling coefficient of $\lambda = 5 \times 10^{-3}$
and a desired acceleration noise of
$4 \times 10^{-13} \ \mathrm{m/sec^2 Hz^{1/2}}$, then we must determine
$\lambda$ to a relative accuracy of 0.003 for the Earth geodesy DMA.
Likewise, from Figure \ref{fig:SRP}, the solar radiation pressure
acceleration noise around 1 mHz is
$1.6 \times 10^{-10} \ \mathrm{m/sec^2 Hz^{1/2}}$.
Again, using $\lambda = 5 \times 10^{-3}$
and a desired acceleration noise of
$3 \times 10^{-15} \ \mathrm{m/sec^2 Hz^{1/2}}$, we must determine
$\lambda$ also to a relative accuracy of 0.003 for the
gravitational wave DMA.

There does exist a fundamental limit to how well these cross-coupling
forces can be determined and subtracted in the analysis.
We assume that the best possible voltage reference
is limited to a relative voltage stability of $2\times10^{-6}$.
For the geodesy application it is reasonable to assume that maximum
cross-track acceleration, which occurs with a polar Earth orbit is $\sim 10^{-6}
\ \mathrm{m/sec^2}$.
Therefore, it is also reasonable to assume that the maximum dynamic range
of the cross track suspension force results in an acceleration that is ten
times this value, or $10^{-5} \ \mathrm{m/sec^2}$.
Finally, assuming a cross coupling coefficient of $5\times10^{-3}$, resulting
acceleration in the along track (sensitive direction) is,
\begin{equation}
  a_x = (10^{-5})(2\times10^{-6})(5\times10^{-3}) = 10^{-13} \ \mathrm{m/sec^2},
  \nonumber
\end{equation}
which is below the fundamental limit shown in Figure \ref{fig:GRACEaccelDM}.

For the the gravitational wave application, again we assume the a 500 kg
LISA-like spacecraft, with cross sectional area of $4 \ \mathrm{m^2}$ is
1 AU from the Sun.
The resulting nominal solar radiation pressure is $4\times10^{-8}
\ \mathrm{m/sec^2Hz^{1/2}}$.
Therefore, if we assume that the maximum required test mass suspension force
is $4\times10^{-7} \ \mathrm{m/sec^2Hz^{1/2}}$, the fundamental
cross-coupling acceleration noise limit is,
\begin{equation}
  a_x = (4\times10^{-7})(2\times10^{-6})(5\times10^{-3}) = 
  4\times10^{-15} \ \mathrm{m/sec^2},
  \nonumber
\end{equation}
which is roughly
equal to the LISA acceleration noise requirement at low frequency.

\section{Measurement noise}
\label{sec:measNoise}

In a DMA we
use a laser interferometer to measure the acceleration of a reference point
on the spacecraft (an optical bench) relative to the test mass, which
we assume is inertially fixed.
Therefore, in addition to the acceleration noise acting on the TM, we
must also consider the acceleration measurement noise of the interferometer.
For the discussions here, we will assume that the interferometer
exhibits a flat amplitude spectral density.
We analyze the position measurement provided by the interferometer between
kicks to estimate the acceleration of the spacecraft.
There are several approaches that can be used, including second order
finite differencing.
One of the best approaches is to fit a parabola to the sampled position data
between kicks.
We fit the following model to the measured data $z(t)$:
\begin{equation}
  z(t) = x_0 + v_0 \, (t - t_0) + \frac{1}{2} \, a_0 \, (t - t_0)^2
\end{equation}
The fit parameters are $x_0$, the mean position, $v_0$, the mean velocity,
and $a_0$, the mean acceleration, which is what we wish to estimate.
This approach, which has the advantage of being linear and
using all of the measured data,
provides one acceleration measurement per kick period,
$T_{\mathrm{kick}}$.
The resulting acceleration measurement noise (standard deviation), $\sigma_a$,
depends linearly on the
interferometer noise level $\sigma_I$,
quadratically on $T_{\mathrm{kick}}^{-1}$,
and inversely on the square root of the number of samples, $N$.
If we assume a constant sampling frequency, say 10 Hz, and a small
but constant duty cycle, say 0.1, then the 
number of samples $N$ is roughly proportional to $T_{\mathrm{kick}}$.
We then have the following relationship between acceleration measurement noise,
interferometer noise and kick period:
\begin{equation}
  \sigma_a \approx \alpha \, \frac{ \sigma_I }{ T_{\mathrm{kick}}^{5/2} }
\end{equation}
The parameter $\alpha$, of order 1, depends on the cross correlation between
the mean acceleration $a_0$ and the constant and linear terms $x_0$ and $v_0$.
Larger kick periods greatly decrease the acceleration measurement noise,
but also greatly increase the maximum displacement of the test mass
relative to its housing.
Larger kick periods also proportionally reduce the bandwidth of the measurement
since one acceleration noise measurement is made every $T_{\mathrm{kick}}$.

Assuming that the interferometer exhibits a white noise spectrum in
displacement, then the
acceleration measurement noise also has a white spectrum
(a linear function of a Gaussian is a Gaussian).
This is one disadvantage of the DMA since the measurement noise spectrum is
flat in acceleration, while a continuous test mass
displacement measurement, uninterrupted by kicks (e.g. using drag-free),
which is then twice differentiated has a $1/f^2$ spectrum in acceleration.
Therefore, the measurement noise in acceleration for a drag-free systems
is much lower at lower frequencies where most of the interesting science is,
assuming a given interferometer noise level.

\begin{figure}
  \centering
  \includegraphics[width = 12 cm]{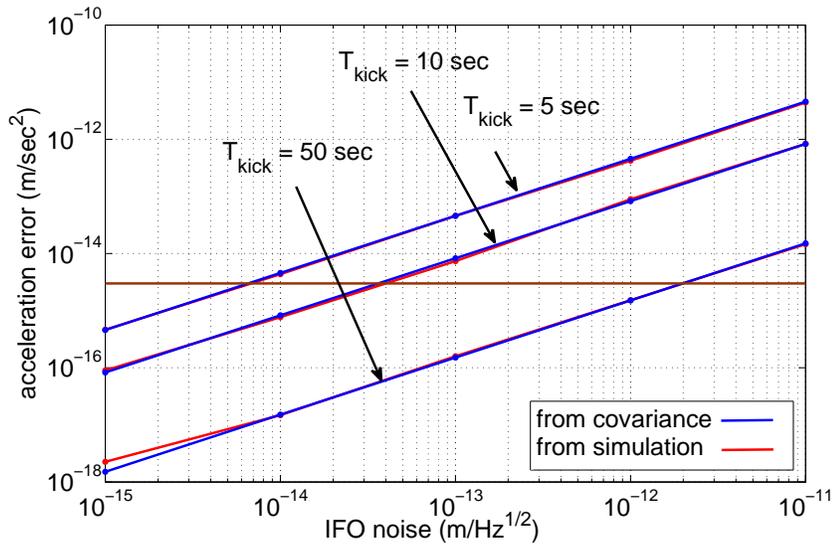}
  \caption{Acceleration measurement noise as a function
   of the kicking period and interferometer noise. \label{fig:measNoise}}
\end{figure}

Figure \ref{fig:measNoise} plots the relationship between acceleration
measurement noise and
interferometer noise for $T_{\mathrm{kick}} = 5 \ \mathrm{sec},
10 \ \mathrm{sec}$, and $50 \ \mathrm{sec}$.
The estimated acceleration measurement noise calculated using the standard
covariance analysis is shown in blue, while red curves show the measurement
error obtain through a numerical simulation.
The simulation assumed a spacecraft acceleration due to the solar radiation
pressure model discussed above, a 0.1 duty cycle, and a sampling rate
of 10 Hz.
The interferometer (IFO) noise was assumed to be white with a standard
deviation as shown on the plot after averaging over 1 sec (10 samples).
We see from Figure \ref{fig:measNoise} that the solar radiation pressure noise
at high frequencies does not adversely affect the acceleration measurement.
For a desired acceleration measurement noise of $3 \times 10^{-15}
 \ \mathrm{m/sec^2 Hz^{1/2}}$ and a kicking period of 10 sec
an interferometer with a white noise level of 40 $\mathrm{fm/Hz^{1/2}}$
is needed.
For a 50 sec kicking period a 2 $\mathrm{pm/Hz^{1/2}}$ interferometer is
needed.

These interferometer requirements only apply to the local (short-arm)
interferometer, which is far from being shot noise limited,
and not the intra-spacecraft (long-arm) interferometer.
In addition, these noise requirements only apply at frequencies above
$(1/T_{\mathrm{kick}}) = 0.1 \ \mathrm{Hz}$ in the case $T_{\mathrm{kick}} =
10 \ \mathrm{sec}$.
Therefore, we need not worry about challenging
low frequency measurement noise, for example due to temperature changes
and thermal expansion or index of refraction changes of materials.
Each short-arm interferometer measurement lasting $T_{\mathrm{kick}}$
seconds is independent of all others.

\section{DMA electrode geometry}

\begin{figure}
  \centering
  \includegraphics[width = 10 cm]{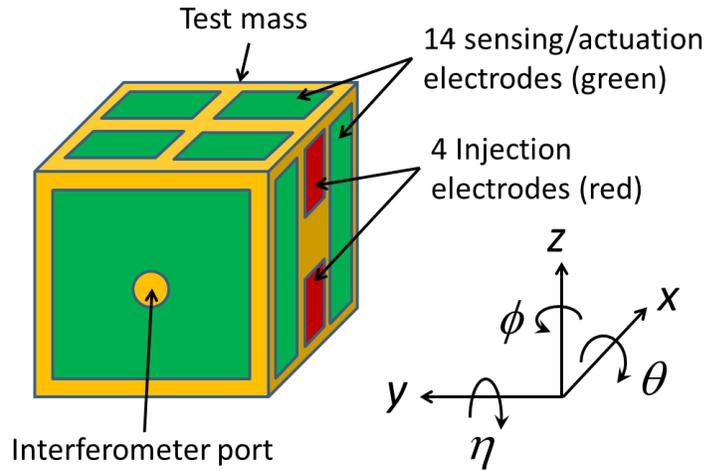}
  \caption{Proposed electrode geometry for the Drift Mode
    Accelerometer. \label{fig:DMAelec}}
\end{figure}

Figure \ref{fig:DMAelec} shows a proposed electrode geometry that is
slightly modified from that of the
LISA Pathfinder GRS \cite{strayFields2012}.
The geometry shown in Figure \ref{fig:DMAelec} maximizes the actuation
authority along the sensitive $x$-axis and at the same time
decouples $x$-axis actuation from that of all other degrees of freedom.
This allows a clean separation of drift-mode operation along $x$ and
continuous suspension in all other degrees of freedom.
A small port is needed in the middle of the $x$-axis electrode to allow
for the interferometric readout along $x$.
Mechanical pins required to cage the test mass during launch would be
located between the two injection electrodes along the $y$-axis.

\section{Testing drift mode accelerometry}

Precision torsion pendula thus far represent the best method of testing the
performance of precision inertial instruments in the laboratory
\cite{trentoPendulum}.
One such pendulum at the University of Florida
consists of a cross bar supported by a 1 m long, 50 $\mu m$ diameter W fiber.
A light-weighted aluminum cubic test mass is mounted at each of the four
ends of the cross bar.
Two electrode housings surround two opposing test masses.
The cross bar is used to convert the rotational motion of the torsion
pendulum into mostly translational motion of the four test masses.
The electrode housings can both electrostatically force the test masses
and readout their position capacitively.
A small port is also incorporated into the electrode housings to allow
for an interferometric readout of the test mass' position.
The entire apparatus is housed in a vacuum chamber.

In order to test the performance of the DMA, the neutral orientation
of the pendulum can be biased so that the pendulum restoring force can
be made equivalent to the dc acceleration of the spacecraft
either due to atmospheric drag or solar radiation pressure.
The electrostatic actuation system can be operated with a low duty cycle
just as described above and a laser interferometer can be used to estimate
the test mass' acceleration.
higher frequency spacecraft disturbances can be simulated by varying the
neutral orientation of the pendulum or by applying noise voltages to the
electrodes that are equivalent to the spacecraft acceleration noise.
With this approach the acceleration noise floor can be measured
and compared with the acceleration noise floor obtained with
the actuation turned off and the pendulum in its neutral orientation
set with the test masses centered in their housings.

The best way to determining the performance of the DMA would be
to test the instrument in space.
The LISA Pathfinder mission offers one opportunity to do this.
If the drag-free and micropropulsion systems were turned off and both
test masses were operated in a drift mode, then the resulting differential
acceleration noise between the two test masses could be estimated using the
on-board laser interferometer.
All cross couplings and stiffness can be determined and accounted for
in the analysis of the data.

\section{Conclusion}

The drift mode accelerometer is a modified electrostatic accelerometer
potentially capable of acceleration noise performance
similar to that of drag-free
systems without the need for drag-free control or associated precision
propulsion.
A DMA consists of a dense test mass that is freely floating inside an
electrode housing, which can both sense its position capacitively and actuate
it electrostatically.
Unlike traditional electrostatic accelerometers, the suspension system
is operated with a low duty cycle and with a cycling frequency that is
chosen to be above the science signals of interest.
Measurement of spacecraft acceleration is made using a laser interferometer,
which is not limited by dynamic range.
Two applications of the DMA, Earth geodesy and
gravitational wave observation, are studied.
Both represent gravitational
science missions where the DMA might be used to replace drag-free operation.
For gravitational wave observation, the combination of the existing
LISA Pathfinder gravitational reference sensor and the LISA
local (short-arm) interferometer 
can be operated as a drift mode accelerometer, with acceleration noise
performance close to that required for LISA.
Detailed modeling and analysis is still required to
fully determine the acceleration noise performance and instrument
requirements and constraints.
Laboratory testing
using torsion pendula provide one promising approach for demonstrating the
performance and operation of the drift mode accelerometer.

\section{Acknowledgments}

The author would like to thank Guido M\"uller and Giacomo Ciani at the
University of Florida and William Weber at the University of Trento
for their valuable insights related to this work.

\end{document}